\documentclass[12pt]{iopart} 

\usepackage{epsfig,multicol,iopams,graphicx}
\newcommand\fverb{\setbox\pippobox=\hbox\bgroup\verb}
\newcommand\fverbdo{\egroup\medskip\noindent%
            \fbox{\unhbox\pippobox}\ }
\newcommand\fverbit{\egroup\item[\fbox{\unhbox\pippobox}]}
\newbox\pippobox

\def\NP{Nucl. Phys. }
\def\PL{Phys. Lett. }
\def\PR{Phys. Rev. }

\def\PRL{Phys. Rev. Lett.}

\def\CMP{Commun. Math. Phys.}
\def\JHEP{J. High Energy Phys.}

\def\Pre{Preprint}

\begin{document}

\title{Hawking radiation from covariant anomalies in 2+1 dimensional black holes}

\author{Soonkeon Nam$^{1,2}$ and Jong-Dae Park$^{3}$}

\address{$^{1}$ Department of Physics and Research Institute of Basic
    Science, \\
    Kyung Hee University, Seoul 130-701, Korea\footnote{Permanent Address}}
\address{$^{2}$ Center for Theoretical Physics, MIT, Cambridge, MA 02139, U.S.A.}
\address{$^{3}$ Center for Quantum Spacetime Sogang University, Seoul
    121-742, Korea}
    \ead{\mailto{nam@khu.ac.kr}, \mailto{jdpark@khu.ac.kr}}


\begin{abstract}
In an insightful approach, Robinson and Wilczek proposed that
Hawking radiation can be obtained as the compensation of a breakdown
of general covariance and gauge invariance and the radiation is a
black body radiation at Hawking temperature. We apply this method to
two types of black holes in three dimensional spacetime, which share
the general form of the metric in that the $tt$ component of the
metric is not inverse of the $rr$ component of the metric. First one
is the warped $AdS_3$ black hole in three dimensional topologically
massive gravity with negative cosmological constant, and second one
is the charged rotating black hole in three dimensions.
\end{abstract}
%
%
\maketitle




\section{Introduction}
Hawking radiation is one of the most important quantum effect when
we are studying the quantum gravity in a black hole background
spacetime.
Classically a black hole does not radiate, but when we consider
quantum effects around a generic black hole geometry we have
radiation. Hawking's original work \cite{Hawking75} have shown that
the calculation of the Bogoliubov coefficients gives a Planck
distribution  of radiation as one of thermodynamical properties.
There are other derivation of Hawking radiation taking account of
the quantum effect in the black hole background as a tunneling
process based on particles in a dynamical geometry
\cite{Hawking77}\cite{Wilczek00}. It is found that Hawking radiation
is determined by the horizon
properties.  
Another approach to understand Hawking radiation is the calculation
of the energy-momentum tensor in black hole backgrounds. The
energy-momentum tensor should be conserved in a curved background in
classical field theory with general covariance.  However, quantum
mechanically there is breaking of symmetry which gives rise to the
anomaly term.

Recently, Robinson and Wilczek \cite{Wilczek05} and also Iso, Umetsu
and Wilczek \cite{Wilczek06} proposed another derivation of the
energy-momentum tensor which describes the Hawking radiation as the
compensation of the anomaly to break the classical symmetry. There
are two types of anomalous currents, i.e. the consistent and the
covariant. The anomaly equations with the consistent currents
satisfy the Wess Zumino consistency condition but do not  transform
covariantly under a gauge transformation. On the other hand,
Banerjee and Kulkarni \cite{Kulkarni07}\cite{Kulkarni07-1} made use
of another method involving covariant current which transforms under
gauge transformation, but which does not satisfy the Wess Zumino
consistency condition. This method  was extended to the case charged
black holes, rotating black holes and other types of black holes
\cite{Wilczek06}\cite{Wilczek06-1}
$\sim$\cite{Chinese}.

In this paper, we calculate the energy-momentum flux of the warped
$AdS_3$ black hole in three dimensional topologically massive
gravity with negative cosmological constant.  In three dimensions
ordinary Einstein gravity is trivial, but with the gravitational
Chern-Simons term there is a propagating massive graviton mode
\cite{Deser82}. In light of this, recently there has been some
development to understand topologically massive gravity with
negative cosmological constant $-1/\ell^2$
\cite{MIPark}$\sim$\cite{ref2}. We also consider another type of
black hole in three dimensions, that of the charged rotating black
hole.

In three dimensional topological massive gravity, there is an
$AdS_3$ vacuum solution for any coefficient $\mu$ of the
Chern-Simons action term. Recently, it was shown that these
solutions are unstable except at the chiral point $\mu\ell =1
~$\cite{Strominger08}. More recently, for $\mu\ell \neq 3$ two other
types of black hole solutions which have $SL(2,\mathbb{R}) \times
U(1)$-invariant warped $AdS_3$ geometries were with timelike or
spacelike $U(1)$ isometry, and at $\mu\ell = 3$ there are also a
pair of solutions with a null $U(1)$ isometry \cite{Strom08}.

It is known that for $\mu\ell > 3$ warped black hole solutions are
asymptotic to warped $AdS_3$. It is also shown that these black hole
solutions can be regarded as discrete quotients of warped $AdS_3$
just like BTZ black holes are discrete quotients of $AdS_3$. Guica
et al.\cite{Strom-0809} have argued that in the very near horizon
region of an extreme Kerr black hole, this geometry becomes warped
$AdS_3$ with an $SL(2,\mathbb{R})\times U(1)$ isometries
\cite{Horo99} and the geometry of a three dimensional slice of fixed
polar angle $\theta$ gives a quotient of warped $AdS_3$ with the
identification of azimuthal angle $\varphi$
\cite{Detournay05}\cite{Bengtsson05}. Such quotients are warped
$AdS$ black holes. With consistent boundary conditions  the
asymptotic symmetry generators give one copy of Virasoro algebra
with a chiral half of central charge. So they have conjectured that
extreme Kerr black holes are dual to a chiral conformal field
theory. At $\mu\ell = 3$ null warped $AdS_3$ is given as a solution
and it can be related to the dual theory between gravity and
conformal quantum mechanical theory \cite{Malda08}.

Here, we will concentrate on a type of spacelike stretched warped
black hole solution \cite{Strom08}. When we want to apply Robinson
and Wilczek's method to the warped stretched $AdS_3$ black hole
which has the different metric form for `$tt$' component and the
inverse of `$rr$' component.

There is another type of three dimensional black hole, which
is charge rotating black hole \cite{Martinez00}.
 In fact, Gangopadhyay and Kulkarni \cite{Kulkarni07-2} have
calculated the energy-momentum flux for a black hole with
$g_{rr} g_{tt} \neq 1$ in a different context.
Their method can be applied to these black holes.


\section{Hawking radiation and Covariant anomalies}
In this section we will review how the Hawking radiation can be
obtained from the covariant anomaly. Consider the
Reissner-Nordstr\"{o}m black hole metric of the following form :
\begin{eqnarray}
 ds^2 = -f(r)dt^2 + \frac{dr^2}{f(r)} + r^2 d\Omega_{d-2}^2 \,.
\end{eqnarray}
Robinson and Wilczek \cite{Wilczek05} have considered that the energy
flux of the Hawking radiation can be given by the effective two dimensional theory
having gravitational anomalies at the black hole horizon.
This was further generalized for the case of charged black holes
where gauge anomaly was also considered for the flux of charge \cite{Wilczek06}.
We can also consider anomaly analysis of Hawking radiation for more general cases \cite{Kulkarni07-2}
of the following type:
\begin{eqnarray}\label{metric1}
 ds^2 = - h(r) dt^2 + \frac{dr^2}{f(r)} + r^2 (d\phi
        + N^{\phi}(r)dt)^2,
\end{eqnarray}
where $\phi$ is an angle $\phi \in [0, 2\pi]$ and the horizon is
located at $r=r_H$ satisfying $f(r_H)=0$ and $h(r_H)=0$.

To consider Hawking radiation via  anomaly, let us first
consider the scalar field on this metric background.
The action of this scalar field is
\begin{eqnarray}
 S[\psi] &=& -\frac{1}{2} \int d^3x \sqrt{-g}
             g^{\mu\nu} \partial_{\mu} \psi \partial_{\nu} \psi \\
         &=& -\frac{1}{2} \int dt dr d\phi \sqrt{\frac{h}{f}}r
             \left[
                \frac{1}{h}\psi \partial^2_t \psi
                - \sqrt{\frac{f}{h}}\frac{1}{r} \psi \partial_r
                (\sqrt{f h}r\partial_r \psi) \right.  \\
         && \left. - 2\frac{N^{\phi}}{h} \psi
                \partial_t\partial_{\phi} \psi
                + \frac{r^2 (N^{\phi} )^2-h}{r^2 h} \psi
                \partial_{\phi}^2 \psi
             \right].
\end{eqnarray}
Performing the partial wave decomposition, $\psi(t,r,\phi) = \sum_m \varphi_m (t,r)
e^{im\phi}$, this action is reduced to an infinite set of the scalar
fields on a two dimensional space. If we go near the horizon,  $r\rightarrow r_H$,
the above action near the horizon can be effectively described with the following form
\begin{eqnarray}
 S[\varphi] = -\frac{1}{2} \sum_m \int dt dr ~ \varphi_m^*
       \left[
          \frac{r}{\sqrt{f h}} (\partial_t - i m N^{\phi})^2
          - \partial_r \sqrt{f h}r \partial_r
       \right] \varphi_m \,.
\end{eqnarray}
From this action, $\varphi_m(t,r)$ as a two dimensional
complex scalar field in the background $U(1)$ static gauge field $A_t (r)=
N^{\phi}(r)$ with charge $m$. When we consider the charged rotating
black hole in three dimensions, this two dimensional effective
theory has two types of $U(1)$ charges. One is from gauge field and
the other from rotation \cite{Wilczek06-1}\cite{Setare06}. These two
types of currents are identical in nature, so we consider one type
of current and then we can use the same result.

The idea of Robinson and Wilczek is to divide the region outside the
horizon of black hole into two. First is the near horizon and the
other sufficiently far away. Near horizon region (H) is described by
two dimensional effective field theory. The radial direction $r$ and
the time $t$ are the two dimensional coordinates. We can put the
outgoing modes near the horizon as right moving modes and ingoing
modes as left moving modes. The horizon being a null hypersurface,
the ingoing modes at the horizon does not affect physics outside the
horizon. It is a chiral theory and therefore there is chiral
anomaly. The region away from the horizon (O) is four dimensional
and is free of anomaly.

The covariant gauge current is anomalous because modes interior to
the horizon can not affect physics outside the horizon. The
covariant gauge current conservation is given by
\begin{eqnarray}
 \nabla_{\mu} \tilde{J}^{\mu} = - \frac{e^2}{4\pi}
 \varepsilon^{\mu\nu} F_{\mu\nu} = \frac{e^2}{2\pi \sqrt{-g}}
 F_{rt},
\end{eqnarray}
where we take $\varepsilon^{\mu\nu} = \epsilon^{\mu\nu}/\sqrt{-g},
\varepsilon_{\mu\nu} = \sqrt{-g}\epsilon_{\mu\nu}$ and
$\epsilon^{tr} = 1 = -\epsilon_{tr}$. The current outside the
horizon ($r\in [r_H+\epsilon,\infty]$) is anomaly free and satisfies the conservation law
\begin{eqnarray}
 \nabla_{\mu} \tilde{J}^{\mu}_{(O)} = 0.
\end{eqnarray}
Near the horizon ($r\in [r_H,r_H+\epsilon]$), there are only outgoing fields and the current
satisfies the anomalous equation
\begin{eqnarray}
 \nabla_{\mu} \tilde{J}^{\mu}_{(H)}
     = \frac{1}{\sqrt{-g}} \partial_r
       ( \sqrt{-g} \tilde{J}^r_{(H)} )
     = \frac{e^2}{2\pi \sqrt{-g}} F_{rt}
     = \frac{e^2}{2\pi \sqrt{-g}} \partial_r A_t \,.
\end{eqnarray}
It appears that we have introduced a very sharp boundary between two
regions (H) and (O). Such a sharp boundary brings in delta functions
in the analysis. There has been some works \cite{Umetsu08} which did
away with such delta functions, but in this paper we follow the
original approach.

Solving above two equations, we get
\begin{eqnarray}
 \sqrt{-g}\tilde{J}^r_{(O)} &=& c_{O},  \label{coeffO}\\
 \sqrt{-g}\tilde{J}^r_{(H)} &=& c_H + \frac{e^2}{2\pi}
           [A_t (r)-A_{t}(r_H)], \label{coeffH}
\end{eqnarray}
where $c_O$ and $c_H$ are integration constants.
Writing the current $\tilde{J}^r$ as a sum of two regions
\begin{eqnarray}
 \tilde{J}^r = \tilde{J}^r_{(O)} \Theta (r-r_H - \epsilon)
      + \tilde{J}^r_{(H)} H(r),
\end{eqnarray}
where $\Theta(r)$ and $H(r) = 1-\Theta(r-r_H -\epsilon)$ are step
functions that are equal to 1 in region (O) and (H) respectively and
vanish elsewhere.

Then by using the conservation equations, the Ward identity becomes
\begin{eqnarray}
 \nabla_{\mu} \tilde{J}^{\mu}
      = ( \tilde{J}^r_{(O)} - \tilde{J}^r_{(H)}
        + \frac{e^2}{2\pi \sqrt{-g}} A_t) \delta(r-r_H -\epsilon)
        + \frac{1}{\sqrt{-g}} \partial_r \left( \frac{e^2}{2\pi}A_t H \right).
\end{eqnarray}
The last term in the total derivative must be canceled by quantum
effects of the classically irrelevant ingoing modes. This quantum
effect induces the Wess-Zumino term by the ingoing modes near the
horizon. Under gauge transformation, the vanishing of the Ward
identity requires that the coefficient of the delta function is
zero. This gives the following condition:
\begin{eqnarray}
 c_O = c_H - \frac{e^2}{2\pi}A_t(r_H) \,.
\end{eqnarray}
The covariant current $\tilde{J}^r_{(H)}$ vanishes at the horizon. So
the coefficient $c_H$ from (\ref{coeffH}) is zero. Therefore the
value of the charge flux is given by
\begin{eqnarray}
 c_O = - \frac{e^2}{2\pi}A_t(r_H) \,.
\end{eqnarray}
If we consider another $U(1)$ charge $m$, then we have two $U(1)$
gauge symmetries and currents. The gauge potential $\mathfrak{B}_t$
should be a sum of two gauge fields $A_t$ and $N^{\phi}$,
\begin{eqnarray}\label{gauge-gen}
 \mathfrak{B}_t = e A_t + m N^{\phi} \,.
\end{eqnarray}
With these $U(1)$ gauge currents, the anomalous equation can be
described by the following form \cite{Wilczek06-1}:
\begin{eqnarray}
 \nabla_{\mu} {\cal J}^{\mu} = \frac{1}{2\pi\sqrt{-g}}
 \partial_r \mathfrak{B}_t \,,
\end{eqnarray}
where ${\cal J}^{\mu}$ is the sum of $U(1)$ currents
$\tilde{ J}^{\mu}_i/e_i~(i=1,2)$ associated with the $U(1)$ charges
$e_1=e, ~e_2=m$ respectively.
In the presence of gauge fields, the energy-momentum tensor does not
preserve the current conservation law but appears the Lorentz force
law. So the corresponding anomalous Ward identity is given by
\begin{eqnarray} \label{gravi-anom}
 \nabla_{\mu} \tilde{T}^{\mu}_{~\nu}
     = {\cal F}_{\mu\nu} {\cal J}^{\mu} + \tilde{\cal A}_{\nu}
     = {\cal F}_{\mu\nu} {\cal J}^{\mu} - \frac{1}{96\pi}
        \varepsilon_{\mu\nu} \partial^{\mu}R \,,
\end{eqnarray}
where  $\tilde{\cal A}_{\nu}$ is the covariant gravitational anomaly.
For the metric (\ref{metric1}), the covariant anomaly is purely
time-like
\begin{eqnarray}
 \tilde{\cal A}_r &=& 0 \,, \\
 \tilde{\cal A}_t
     &=& \frac{1}{\sqrt{-g}}\partial_r \tilde{\cal N}^r_{~t} \,,
\end{eqnarray}
where
\begin{eqnarray} \label{anomaly}
 \tilde{\cal N}^r_{~t}
     = \frac{1}{96\pi} \left( \frac{1}{2}f'h'+f h''
       -\frac{fh'^2}{h} \right) \,.
\end{eqnarray}
Outside region of the horizon, the energy-momentum tensor
satisfies the conservation law
\begin{eqnarray}
 \nabla_{\mu} \tilde{T}^{\mu}_{(O)\nu}
       = {\cal F}_{\mu\nu} {\cal J}^{\mu} \,.
\end{eqnarray}
Considering the Ward identity for the component $\nu=t$, the
conservation equation gives
\begin{eqnarray}
 \nabla_{\mu} \tilde{T}^{\mu}_{~t}
   = \partial_r \tilde{T}^r_{~t}
     + \partial_r(\ln\sqrt{-g})\tilde{T}^r_{~t}
   = \frac{1}{\sqrt{-g}} \partial_r (\sqrt{-g}\tilde{T}^r_{(O)t})
   = {\cal F}_{rt} {\cal J}^r_{(O)}.
\end{eqnarray}
Using (\ref{coeffO}) with the change from $\tilde{J}^{\mu}_i$ to
${\cal J}^{\mu}$,  we get
\begin{eqnarray}\label{em-out}
 \sqrt{-g} \tilde{T}^r_{(O)t} = a_O + c_O \mathfrak{B}_t(r),
\end{eqnarray}
where $a_O$ is an integration constant.
Near the horizon, the covariant energy-momentum tensor has the
gravitational anomaly \cite{Kulkarni07} and this anomaly equation
reads from equation (\ref{gravi-anom})
\begin{eqnarray}
 \nabla_{\mu} \tilde{T}^{\mu}_{(H)t}
     = \frac{1}{\sqrt{-g}} \partial_r(\sqrt{-g}\tilde{T}^r_{(H)t})
     = {\cal F}_{rt} {\cal J}^r_{(H)}
       + \frac{1}{\sqrt{-g}} \partial_r \tilde{\cal N}^r_{~t} \,.
\end{eqnarray}
Using ${\cal J}^r_{(H)}$ from equation (\ref{coeffH}) changed from
$\tilde{J}^r_{(H)}$ to ${\cal J}^r_{(H)}$ and from $A_t(r)$ to
$\mathfrak{B}_t(r)$, the above equation solves
\begin{eqnarray}\label{em-horizon}
 \sqrt{-g}\tilde{T}^r_{(H)t} = a_H + \int^r_{r_H} dr \partial_r
    [c_O \mathfrak{B}_t(r) + \frac{1}{4\pi} \mathfrak{B}^2_t(r)
    + \tilde{\cal N}^r_{~t}(r)] ,
\end{eqnarray}
where $a_H$ is an integration constant. Writing the energy-momentum
tensor as a sum of two contributions
\begin{eqnarray}
 \tilde{T}^{\mu}_{~\nu} = \tilde{T}^{\mu}_{(O)\nu} \Theta(r-r_H - \epsilon)
    + \tilde{T}^{\mu}_{(H)\nu}H(r),
\end{eqnarray}
we find the following form for the covariant derivative of the
energy-momentum tensor:
\begin{eqnarray}
 \nabla_{\mu} \tilde{T}^{\mu}_{~t}
    &=& \frac{c_O}{\sqrt{-g}} \partial_r \mathfrak{B}_t
        + \frac{1}{\sqrt{-g}}\partial_r
          \left(
             \frac{1}{4\pi} \mathfrak{B}_t^2(r) H(r)
             + \tilde{\cal N}^r_{~t}(r) H(r)
          \right)  \\
    &+& \left(
             \tilde{T}^r_{(O)t} - \tilde{T}^r_{(H)t}
             + \frac{1}{4\pi\sqrt{-g}} \mathfrak{B}_t^2(r)
             + \frac{1}{\sqrt{-g}} \tilde{\cal N}^r_{~t}(r)
          \right) \delta(r-r_H -\epsilon).
\end{eqnarray}
The first term is a classical effect of the Lorentz force. The
second term should be canceled by the quantum effect of the
irrelevant ingoing modes. The quantum effect to cancel this term is
the Wess-Zumino term induced by the ingoing modes near the horizon.
Under the diffeomorphism invariance at the horizon, the last term
should  vanish
\begin{eqnarray}
 \tilde{T}^r_{(O)t}
         = \tilde{T}^r_{(H)t} - \frac{1}{4\pi\sqrt{-g}} \mathfrak{B}_t^2
         - \frac{1}{\sqrt{-g}} \tilde{\cal N}^r_{~t} \,.
\end{eqnarray}
Substituting equations (\ref{em-out}) and (\ref{em-horizon}) into
above equation, we obtain
\begin{eqnarray}
 a_O = a_H + \frac{1}{4\pi} \mathfrak{B}_t^2(r_H)
       - \tilde{\cal N}^r_{~t}(r_H) \,.
\end{eqnarray}
The integration constant $a_H$ can be fixed by requiring the
covariant energy-momentum tensor to vanish at the horizon and this
gives $a_H = 0$. Therefore the final result is given by
\begin{eqnarray}
 a_O &=& \frac{1}{4\pi} \mathfrak{B}_t^2(r_H)
         - \tilde{\cal N}^r_{~t}(r_H)  \\
     &=& \frac{1}{4\pi} \mathfrak{B}_t^2(r_H) -
         \left.\frac{1}{96\pi} \left( \frac{1}{2}f'h'+f h''
         -\frac{fh'^2}{h} \right)\right|_{r=r_H}, \label{energy-mom}
\end{eqnarray}
where we  have used equation (\ref{anomaly}). This value can be
interpreted as the energy-momentum flux which coincides with the
flux from black body radiation with a chemical potential at
temperature $T_H$ \cite{Wilczek05}\cite{Wilczek06}. This form of
energy-momentum flux have already been mentioned in
\cite{Kulkarni07-2}. The second term of (\ref{energy-mom}) can be
represented by the Hawking temperature $T_{H}$ which is related to
the surface gravity. Surface gravity can be calculated as follows:
\begin{eqnarray}
 && \kappa^2
    = -\left.\frac{1}{2}(\nabla^{\mu}\xi^{\nu})
        (\nabla_{\mu}\xi_{\nu})\right|_{\cal N}
        = \left.\frac{f}{h} \left(\frac{h'}{2}\right)^2
        \right|_{r=r_H},
 \\
 && \therefore ~~ \kappa = \left. \frac{1}{2} \sqrt{\frac{f}{h}}h'
        \right|_{r=r_H}  \,. \label{surface-gv}
\end{eqnarray}
The relationship between Hawking temperature and surface gravity is
given by $T_H = 1/\beta = \frac{\kappa}{2\pi}$.  Coming back to
equation (\ref{energy-mom}), we focus on the second term. For
simplicity sake we will consider the case when $h(r) = f(r)g(r)$ and
$g(r_H)\neq 0$. In this case the flux becomes
\begin{eqnarray}
a_O &=& \frac{1}{4\pi} \mathfrak{B}_t^2(r_H) +
         \frac{1}{192\pi} g f'^2 \\
      &=& \frac{1}{4\pi} \mathfrak{B}_t^2(r_H) +
         \frac{\pi}{12\beta^2}.
 \label{energy-momnew}
\end{eqnarray}
This form is the key formula in discussions of Hawking radiation.


\section{Hawking radiation of warped $AdS_3$ black hole}
The action of the three dimensional gravity with negative
cosmological constant $\Lambda = -1/l^2$ and gravitational
Chern-Simons term is \cite{Deser02}
\begin{eqnarray}
 S &=& \frac{1}{16\pi G} \int d^3x \sqrt{-g}
       \left( R + \frac{2}{l^2} \right)
       -\frac{1}{32\pi G \mu} \int \Gamma \wedge
       \left( d\Gamma + \frac{2}{3} \Gamma^2 \right) \,.
\end{eqnarray}
The equations of motion are
\begin{eqnarray}\label{eins-eq}
 R_{\mu\nu} - \frac{1}{2}g_{\mu\nu}R - \frac{1}{l^2}g_{\mu\nu}
      = -\frac{1}{\mu}C_{\mu\nu},
\end{eqnarray}
where $C_{\mu\nu}$ is the Cotten tensor defined by
\begin{eqnarray}
 C_{\mu\nu} \equiv \epsilon_{\mu}^{~\rho\sigma} \nabla_{\rho}
     (R_{\sigma\nu} - \frac{1}{4}g_{\sigma\nu}R) \,.
\end{eqnarray}
Clearly all the vacuum solutions of Einstein's equations
automatically satisfy the equation of motion of topologically
massive gravity. Recently, Anninos et al. \cite{Strom08} have found
that there exist two other vacuum solutions which is given by
$SL(2,\mathbb{R}) \times U(1)$ - invariant warped $AdS_3$ geometries
with a timelike or spacelike $U(1)$ isometry for $\mu \ell \neq 3$.
The warping transition from a stretching solution to a squashing one
occurs at critical point $\mu\ell = 3$. Two null warped $AdS_3$
solutions with a null $U(1)$ isometry also exist. There are known
warped black hole solutions which are asymptotic to $AdS_3$ for
$\mu\ell > 3$.
We will only consider the solution for the asymptotically spacelike
stretched case ($\mu\ell > 3$) which are free of naked closed
timelike curves \cite{Clement07}. We can put the coefficient of the Chern-Simons
action as $-\ell/(96\pi G\nu)$ in terms of the dimensionless coupling $\nu=\mu\ell/3$.
Then the metric describing the spacelike stretched black holes for $\nu^2 > 1$ is
given in Schwarzchild coordinates by
\begin{eqnarray}
 \frac{ds^2}{\ell^2} &=& dt^2 + \frac{dr^2}{(\nu^2+3)(r-r_+)(r-r_-)}
     + (2\nu r - \sqrt{r_+ r_- (\nu^2+3)})dt d\theta \nonumber\\
     &+& \frac{r}{4} \left( 3(\nu^2-1)r + (\nu^2+3)(r_+ + r_-)
     - 4\nu \sqrt{r_+ r_- (\nu^2+3)} \right) d\theta^2,
\end{eqnarray}
where $r \in [0, \infty], t \in [-\infty, \infty]$ and $\theta \sim
\theta + 2\pi$. In the ADM formalism, this metric can be written as
\begin{eqnarray}\label{ADM}
 ds^2 = - N(r)^2 dt^2 + \ell^2 R(r)^2 (d\theta + N^{\theta}(r) dt)^2
        + \frac{\ell^4 dr^2}{4R(r)^2 N(r)^2},
\end{eqnarray}
where
\begin{eqnarray}
 R(r)^2 &=& \frac{r}{4} \left( 3(\nu^2-1)r + (\nu^2+3)(r_+ + r_-)
            -4\nu\sqrt{r_+ r_- (\nu^2+3)} \right),
            \label{R-sq}\\
 N(r)^2 &=& \frac{\ell^2(\nu^2+3)(r-r_+)(r-r_-)}{4R(r)^2},  \\
 N^{\theta}(r) &=& \frac{2\nu r - \sqrt{r_+ r_- (\nu^2+3)}}{2R(r)^2}.
   \label{N-th-sq}
\end{eqnarray}
Here  $r=r_+$ and $r=r_-$ are locations of  outer and inner horizons respectively.
In order to obtain energy-momentum flux, we change the above metric to
the form (\ref{metric1}). If we put $\ell R(r)=\rho$, then the
metric (\ref{ADM}) becomes
\begin{eqnarray}
 ds^2 = -N(\rho)^2 dt^2 + \rho^2
        (d\theta + N^{\theta}(\rho)dt)^2
        + \frac{\ell^2 d\rho^2}{N(\rho)^2 (R^2)'^2},
\end{eqnarray}
where $(R^2)'$ is a function of $\rho$ which is described by a
differentiation of $R^2(r)$ with respect to $r$. Comparing this
metric with
\begin{eqnarray}
 ds^2 = -h(\rho) dt^2 + \frac{d\rho^2}{f(\rho)}
        + \rho^2 (d\theta + N^{\theta}(\rho)dt)^2 ,
\end{eqnarray}
we can define
\begin{eqnarray}
 h(\rho) = N(\rho)^2, ~~~ f(\rho) = \frac{(R^2)'^2
 N(\rho)^2}{\ell^2}.
\end{eqnarray}
With these functions, we can find the energy-momentum flux which is
given by (\ref{energy-mom}). The anomaly term at the horizon $\rho_H
= \rho_+ = \ell R(r_+) = \ell R_+$ gives
\begin{eqnarray}
 \tilde{\cal N}^{\rho}_{~t}(\rho_+) &=& \frac{1}{96\pi}
        \left.\left( \frac{1}{2}f'h' + f h''
        - \frac{f h'^2}{h} \right)\right|_{\rho=\rho_+}  \\
        &=& \frac{1}{96\pi} \left.\left( -\frac{2}{\ell^6} \rho^2
            (N^2)'^2 \right)\right|_{\rho=\rho_+}
\end{eqnarray}
where $f', h'$ and $h''$ are functions of $\rho$ which have
differentiations with respect to $\rho$ but $(N^2)'$ gives a
derivative with respect to $r$ as a function of $\rho$.
We have
\begin{eqnarray}\label{N-sq-diff}
 \left.(N(r)^2)'\right|_{r=r_+}
    = \frac{\ell^3(\nu^2+3)}{4\nu}\frac{\rho_+ -\rho_-}{\rho_+^2},
\end{eqnarray}
using the relation between $r$ and $\rho$, i.e. $\ell R(r_{\pm}) =
\rho_{\pm}$.
So the anomaly term becomes
\begin{eqnarray}\label{anom-rho}
 {\cal N}^{\rho}_{~t}(\rho=\rho_+) =
     -\frac{1}{192\pi} \frac{(\nu^2+3)^2}{4\nu^2}
      \frac{(\rho_+ - \rho_-)^2}{\rho_+^2}.
\end{eqnarray}
From (\ref{surface-gv}) we can also find surface gravity
\begin{eqnarray}\label{s-gr-rho}
 \kappa = \frac{1}{2} \left.\sqrt{\frac{f}{h}}h'\right|_{\rho=\rho_+}
    = \frac{1}{\ell^3} \rho \left.(N^2)'\right|_{\rho=\rho_+}
    = \frac{(\nu^2+3)}{4\nu} \frac{\rho_+ - \rho_-}{\rho_+}
    = \frac{2\pi}{\beta} \,.
\end{eqnarray}
Using above two formulas (\ref{anom-rho}) and
(\ref{s-gr-rho}) we have
\begin{eqnarray}
 {\cal N}^{\rho}_{~t}(\rho=\rho_+)
     = -\frac{4}{192\pi} \frac{4\pi^2}{\beta^2}
     = \frac{\pi}{12\beta^2} \,.
\end{eqnarray}
If we substitute $r_{\pm}$ into (\ref{R-sq}), then we can obtain
$R_+^2$ and $R_-^2$ respectively. In Ref.\cite{Strom08}
it is noted that there exist physical black hole so long as $r_+$ and
$r_-$ are non-negative in the parametric region given by $\nu^2 >
1$. With this condition we have
\begin{eqnarray}\label{R-pm}
 R_{\pm} &=& \frac{2\nu r_{\pm} - \sqrt{(\nu^2+3)r_+ r_-}}{2}\,.
\end{eqnarray}
Using this (\ref{R-pm}) we can change the $\rho_{\pm}$ into the
Schwarzschild coordinates $r_{\pm}$. So we get
\begin{eqnarray}
 \frac{\rho_+ - \rho_-}{\rho_+}
     = \frac{R_+ - R_-}{R_+}
     = \frac{2\nu(r_+ - r_-)}{2\nu r_{\pm} - \sqrt{(\nu^2+3)r_+
     r_-}}\,.
\end{eqnarray}
We can now read off the Hawking temperature from the surface gravity
equation.(\ref{s-gr-rho}) $\kappa = 2\pi/\beta = 2\pi T_H$ in
Schwarzschild coordinates and the Hawking temperature is given by
\begin{eqnarray}
 T_H = \frac{(\nu^2+3)(r_+ - r_-)}{4\pi (2\nu r_+ - \sqrt{(\nu^2+3)r_+ r_-})}
 \,.
\end{eqnarray}
The gauge potential $\mathfrak{B}_t$ from (\ref{gauge-gen}) at the
horizon $r_+ = r(\rho_+)$ is given by
\begin{eqnarray}
 \mathfrak{B}_t(\rho_+) &=& n\Omega(\rho_+) = -nN^{\theta}(\rho_+)
 \,.
\end{eqnarray}
Using (\ref{N-th-sq}) and (\ref{R-pm}) we can read off the angular
velocity in Schwarzschild coordinate system
\begin{eqnarray}
N^{\theta}(r_+) = \frac{2}{2\nu r_+ - \sqrt{(\nu^2+3)r_+ r_-}} \,.
\end{eqnarray}
In $(t, \rho, \theta)$ coordinate system, this angular velocity
reduces to $N^{\theta}(\rho_+) = \ell/\rho_+ =1/R_+$.
So, the total flux of the energy-momentum tensor becomes
\begin{eqnarray}
 a_O = \frac{n^2}{4\pi} \frac{\ell^2}{\rho_+^2}
       + \frac{\pi}{12\beta^2}
     = \frac{n^2}{4\pi}
       \left( \frac{2}{2\nu r_+ - \sqrt{(\nu^2+3)r_+ r_-}}
       \right)^2 + \frac{\pi}{12\beta^2} \,.
\end{eqnarray}

At $\nu^2=1$, the metric (\ref{ADM}) reduces to the BTZ black hole in
a rotating frame. If we set $d\phi = d\theta + d\tau/\ell$ and
$d\tau/\ell = dt/(\sqrt{r_+}-\sqrt{r_-})^2$, then we can find the
angular velocity at $r_+$ to be $\Omega_+ = \rho_-/(\ell\rho_+)$.
However, at $\nu^2 > 1$ if we take a similar rotating frame
which has $d\phi = d\theta+\nu r dt/R(r)^2$ then the angular
velocity becomes $\Omega_+ = \sqrt{\rho_+^2-\rho_-^2}/(\ell\rho_+)$.
As $\nu^2$ goes to 1, the warped $AdS_3$ black hole does not go to
BTZ black hole.


\section{Hawking radiation of charged rotating black hole in 2+1-D}
In this section we will consider yet another 2+1 dimensional black hole.
Let us now consider the case of a charged rotating black hole in three
dimensional space-times. Its metric is given by a form
\cite{Martinez00}
\begin{eqnarray} \label{metric-R-BTZ}
 ds^2 = -N^2 F^2 dt^2 + F^{-2} dR^2 + R^2 (d\phi+N^{\phi}dt)^2,
\end{eqnarray}
where
\begin{eqnarray}
 R^2 &=& \frac{r^2-\omega^2 f^2}{1-\omega^2},  \\
 F^2 &=& \left( \frac{d R}{dr} \right)^2 f^2,  \\
 N &=& \frac{r}{R} \left( \frac{dr}{d R} \right)
       = \left( \frac{dr^2}{d R^2} \right),    \\
 N^{\phi} &=& \frac{\omega(f^2 - r^2)}{(1-\omega^2)R^2} \, .
\end{eqnarray}
In order to obtain this rather complicated looking metric, one can``rotation boost" a charged
non-rotating black hole solution.
Consider the following boost along $\phi$ direction, making the black hole to rotate:

\begin{eqnarray}
 \tilde{t} = \frac{t-\omega \phi}{\sqrt{1-\omega^2}}, \ \ \ \
 \tilde{\phi} = \frac{\phi-\omega t}{\sqrt{1-\omega^2}},
\end{eqnarray}
where  $\omega^2 \leq 1$.
The  charged non-rotating black hole solution we start with is of the form
\begin{eqnarray}
 ds^2 = - f^2(r) d\tilde{t}^2 + f^{-2}(r)dr^2 + r^2 d\tilde{\phi}^2,
\end{eqnarray}
where
\begin{eqnarray}
 f^2(r) = r^2 - \tilde{M} - \frac{1}{4}\tilde{Q}^2 \ln r^2.
\end{eqnarray}
Here $\tilde{M}$ and $\tilde{Q}$ are ADM mass and total charge of the black hole.
After the rotation boost, one obtains following forms for the functions in the metric:
\begin{eqnarray}
 R^2 &=& r^2 + \frac{\omega^2}{1-\omega^2}\left( \tilde{M}
         + \frac{\tilde{Q}^2}{4}\ln r^2 \right) \label{r-coordi},\\
 F^2 &=& \frac{\left( r^2
         + \frac{\omega^2 \tilde{Q}^2}{4(1-\omega^2)} \right)^2}{R^2 r^2}
         (r^2 - \tilde{M} - \frac{1}{4} \tilde{Q}^2 \ln r^2)  \label{horizon}, \\
 N &=&  \frac{r^2}{r^2 + \frac{\omega^2 \tilde{Q}^2}{4(1-\omega^2)}},
        \\
 N^{\phi} &=& -\omega \frac{\tilde{M}
          + \frac{1}{4}\tilde{Q}^2 \ln r^2}{(1-\omega^2)R^2} \label{ang-mom} \,.
\end{eqnarray}
The electric potential $A = - \tilde{Q}\ln r d\tilde{t}$ transforms into
a mixture of electric and magnetic potential,
\begin{eqnarray}
 A = -\frac{\tilde{Q}}{\sqrt{1-\omega^2}}\ln r (dt - \omega d\phi)
 \,.
\end{eqnarray}

Considering the metric (\ref{metric-R-BTZ}), we can replace the
function $h(r)$ and $f(r)$ by $N^2(R)F^2(R)$ and $F^2(R)$
respectively. Then we can obtain
\begin{eqnarray}
 \tilde{\cal N}^r_{~t}
      = \left. -\frac{1}{192\pi} N^2 (F^2)'^2 \right|_{R = R_H},
\end{eqnarray}
and surface gravity is
\begin{eqnarray}
 \kappa = \left. \frac{1}{2}N (F^2)'\right|_{R = R_H}
     = \frac{2\pi}{\beta} = 2\pi T_H ,
\end{eqnarray}
where $\beta$ is the inverse of the Hawking temperature $T_H$. The
total flux of energy-momentum tensor is given by \cite{Wilczek06-1}
\begin{eqnarray}
 a_O &=& \frac{1}{4\pi} (e A_t(R_H) + m \Omega(R_H))^2
         + \frac{\pi}{12\beta^2},
\end{eqnarray}
where $\Omega(R_H) = - N^{\phi}(R_H)$. From the equations
(\ref{r-coordi}) and (\ref{horizon}), we can find the event horizon
at
\begin{eqnarray}\label{R-hor}
 R_H^2 = -\frac{\tilde{Q}^2}{4(1-\omega^2)}
         {\cal W}_k \left( -\frac{4}{\tilde{Q}^2}
         e^{-\frac{4}{\tilde{Q}^2} \tilde{M}} \right) ,
\end{eqnarray}
where a subscript of function ${\cal W}_k(x)$, $k=0$ and $-1$.
There has been some work calculating Hawking radiation of (2+1)
dimensional black holes \cite{QQJ}. In that paper, non-rotating but
charged and rotating but chargeless black hole cases were considered
separately. In this paper, since we considered a rotating and
charged black hole, it can be regarded as a unified formulation.

In the above formula, ${\cal W}_k(x)$ is the Lambert function which
is defined as the inverse of the function $f(y)= y e^y$. So we have
\begin{eqnarray}
 && x = y e^y = f(y), \\
 && y = f^{-1}(x) = {\cal W}(x),
\end{eqnarray}
where the integer $k$ denotes the branch of ${\cal W}_k(z)$ when we consider the
extension to complex plane $x \rightarrow z$. ${\cal W}_0(x)$ and
${\cal W}_{-1}(x)$ are the only branches of Lambert function ${\cal
W}$ which has real values.

From (\ref{R-hor}), we find that ${\cal W}_k(x)$ should have
negative values.
${\cal W}_0(x)$ is an increasing function from ${\cal W}_0(-1/e)=-1$
to infinity, passing through the point ${\cal W}_0(0)=0$. On the
other hand ${\cal W}_{-1}(x)$ decreases from ${\cal
W}_{-1}(-1/e)=-1$ to ${\cal W}_{-1}(0^-)=-\infty$. If $\tilde{M}
\rightarrow \infty$, then outer horizon from equation (\ref{R-hor})
goes to infinity and inner horizon becomes 0. So  in general for a
given value of $x$, ${\cal W}_{-1}(x)$ and ${\cal W}_0(x)$
correspond to outer and inner horizon respectively. When the
argument of equation. (4.17) is equal to $x=-1/e$, ${\cal
W}_0(-1/e)$ and ${\cal W}_{-1}(-1/e) $ have the same value of $-1$
and it corresponds to the extremal case where $\tilde{M} =
\tilde{Q}^2 ( 1 - \ln (\tilde{Q}^2/4))/4$. Since the outer horizon
has to stay outside of the inner horizon, the range of the argument
$x$ is $x \in [-1/e, 0]$.

The Lambert function occurs in the solution of time-delayed
differential equation $y'(t)=ay(t-1)$. This solution appears as a
form $y(t)=\exp({\cal W}_k(a)t)$. For more details, one can see in Ref.
\cite{Lambert}. In the following figures we show two Lambert
functions $-{\cal W}_0(x)$ and $-{\cal W}_{-1}(x)$.

\begin{figure}
\begin{center}
  \includegraphics{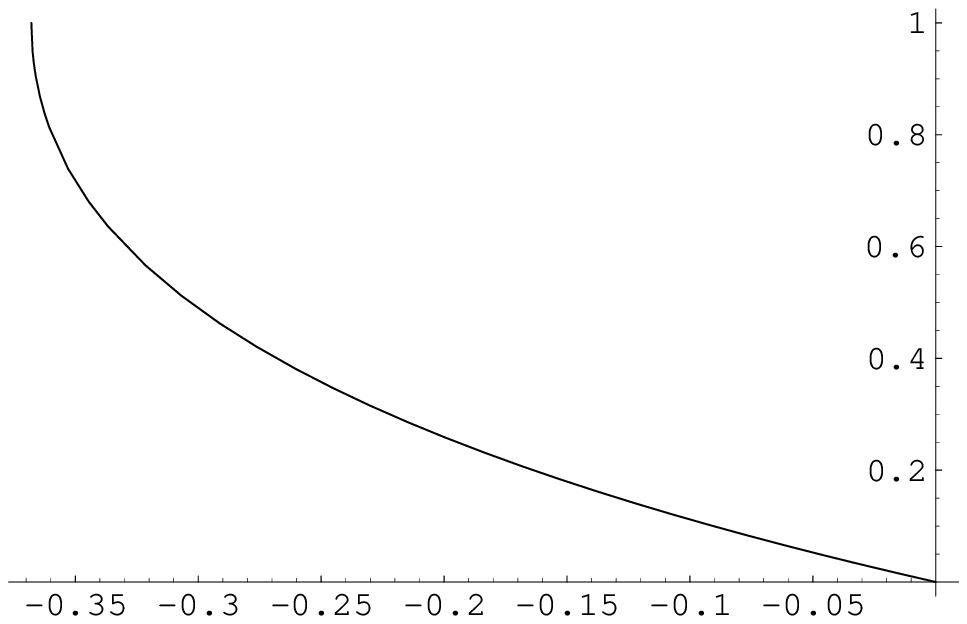}
  \caption{Lambert function : $-{\cal W}_0(x)$}
\end{center}
\begin{center}
  \includegraphics{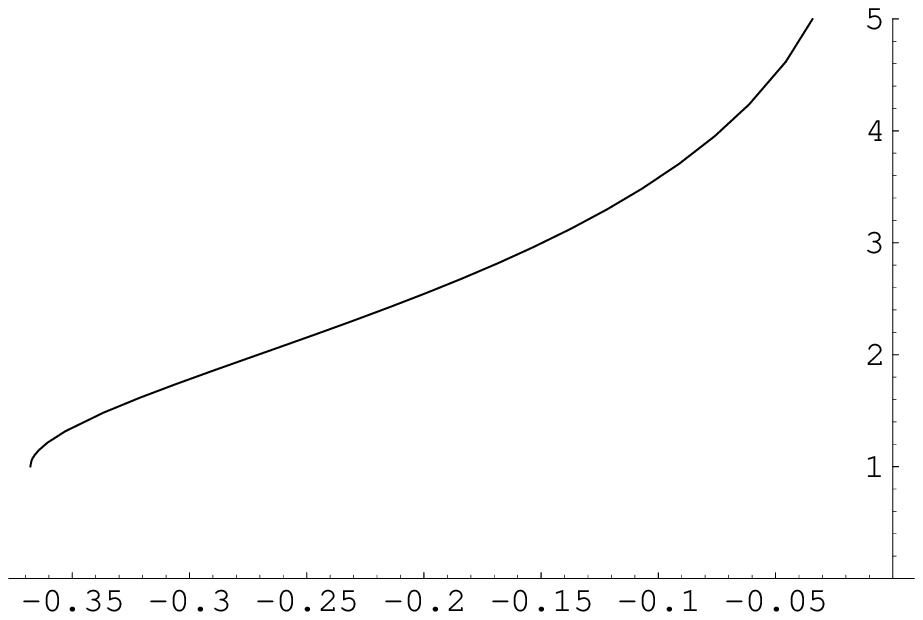}
  \caption{Lambert function : $-{\cal W}_{-1}(x)$}
\end{center}
\end{figure}

From (\ref{r-coordi}), the function $R^2$ as a function of $r^2$ has two
branches separated by a singularity which makes an infinite throat
at $r^2=0$. The region of the black hole corresponds to $R^2 \geq
0$. In the limit $\tilde{Q} \rightarrow 0$, $R^2$ becomes a linear
function of $r^2$ and $R^2=R_-^2 = \omega^2 \tilde{M}/(1-\omega^2),$
which represents the inner horizon of the uncharged rotating black
hole at $r^2=0$.

When the electric charge turned on, this throat
disconnects inner horizon $R_-$ from the black hole space.
 Furthermore, a new inner horizon appears in the black hole space. This
describes the perturbative instability of the inner horizon. In
(\ref{R-hor}), outer and inner horizon are represented by ${\cal
W}_{-1}$ and ${\cal W}_0$ respectively. In order to have two
horizons, the argument of the Lambert function from (\ref{R-hor})
has to satisfy the condition $-1/e < x \leq 0$. This gives the
condition between mass $\tilde{M}$ and charge $\tilde{Q}$,
\begin{eqnarray}
 \tilde{M} > \frac{\tilde{Q}^2}{4}
    \left( 1-\ln\frac{\tilde{Q}^2}{4} \right),
\end{eqnarray}
and we can see the same condition from Ref.\cite{Martinez00}.
In the Fig.1 and Fig.2, there are two points describing values of
the Lambert function ${\cal W}_0$ and ${\cal W}_{-1}$ as inner and
outer horizon respectively. So we can see the outer horizon becomes
\begin{eqnarray}
 R_{H_{out}}^2 = -\frac{\tilde{Q}^2}{4(1-\omega^2)}
                 {\cal W}_{-1} \left( -\frac{4}{\tilde{Q}^2}
                 e^{-4\tilde{M}/\tilde{Q}^2} \right) \,.
\end{eqnarray}
With the above outer horizon, we can obtain electric potential and
angular momentum as follows ;
\begin{eqnarray}
 A_t(R_{H_{out}}) &=& - \frac{\tilde{Q}}{\sqrt{1-\omega^2}} \ln
              (R_{H_{out}} \sqrt{1-\omega^2}) \label{potential},\\
 \Omega(R_{H_{out}}) &=& - N^{\phi}(R_{H_{out}})
              = \omega \label{angular}\,.
\end{eqnarray}
Therefore, the final result  for the flux of Hawking radiation becomes
\begin{eqnarray} \label{flux-ch}
 a_O = \frac{1}{4\pi}
       \left(
          \frac{e\tilde{Q}}{\sqrt{1-\omega^2}}
          \ln(R_{H_{out}} \sqrt{1-\omega^2})
          + m\omega
       \right)^2
       + \frac{\pi}{12\beta^2},
\end{eqnarray}
which is the form of equation (\ref{energy-mom}) with the
$\mathfrak{B}_t$ as a form of (\ref{gauge-gen}).
We have to consider the distribution of fermion in order to avoid superradiance in case of
rotating black holes \cite{Gibbons75}.
The Hawking radiation is given by the Planck distribution with chemical potentials for electric charge $e$ of the
field and an azimuthal angular momentum $m$ from the black hole.
The fermion distribution for this black hole is given by
\begin{eqnarray}
 J_{\pm e,\pm m}(\varepsilon)
    = \frac{1}{e^{\beta(\varepsilon \pm e\Phi \pm m\Omega)}+1},
\end{eqnarray}
where $\Phi$ and $\Omega$ are defined by equation (\ref{potential})
and (\ref{angular}). From these distributions, we can also recover
the same result (\ref{flux-ch})
\begin{eqnarray}
 a_O = \int^{\infty}_{0} \frac{d\varepsilon}{2\pi} \varepsilon
       (J_{e, m}(\varepsilon) + J_{-e,-m}(\varepsilon))
       = \frac{1}{4\pi}(e\Phi+m\Omega)^2
       + \frac{\pi}{12\beta^2}.
\end{eqnarray}
The first term of this value of the flux represents the chemical
potential and the last term means the Hawking flux of this black
hole.
%




%
\section{Summary and outlook}

In this paper, we studied the Hawking radiation from black holes in
three dimensional spacetime using methods of covariant anomaly
cancelation \cite{Kulkarni07} based on works of Robinson and Wilczek
\cite{Wilczek05}. In particular, we have calculated the total flux
of the energy-momentum tensor from warped spacelike stretched
$AdS_3$ black hole in topologically massive gravity with negative
cosmological constant, and also for the case of charged rotating
black hole in three spacetime dimensions. The metric form of these
two kinds of black holes does not have the same form that of the
Reisner-Nordstr\"{o}m black hole. The warped spacelike stretched
$AdS_3$ vacuum solution can be written as a spacelike Hopf fibration
over Lorentzian $AdS_2$ with fiber as a real line. In the case of
the warped spacelike stretched $AdS_3$ black hole, the metric form
(\ref{ADM}) have the same form that of the warped spacelike
stretched $AdS_3$ vacuum solution but a warped factor as a function
of $r$ not constant. The metric form (\ref{metric1}) can be obtained
by the ``rotation boost" \cite{Martinez00}.
The radius of the event horizon can be determined by the solution of
the equation (\ref{horizon}) which can be described using Lambert
function. The outer horizon is described by the Lambert function
with its branch $k=-1$, ${\cal W}_{-1}(x)$ and inner horizon by the
Lambert function with $k=0$, ${\cal W}_0(x)$. The flux of Hawking
radiation (\ref{flux-ch}) is obtained which is dependent on the
outer horizon, in terms of ${\cal W}_{-1}$.  There are many
other higher dimensional black holes \cite{MPradi} and black rings
\cite{Ring} which was studied by this method. However, there are
even more black objects which can in principle be studied by this
method. It would be interesting to further our analysis to this
objects.
\ack{This work is supported by the SRC program of KOSEF through
CQUeST (R11-2005-021) and the Kyung Hee University Research Fund.}




%

\end{document}